\documentclass[aps, prd, twocolumn, showpacs, floatfix, letterpaper,
  nofootinbib, superscriptaddress,longbibliography]{revtex4-1}

\usepackage{graphicx} \usepackage{epsfig} \usepackage{bm}
\usepackage{amsfonts} \usepackage{pgf} \usepackage{color}
\usepackage[T1]{fontenc} \usepackage[latin1]{inputenc}
\usepackage{graphicx} \usepackage[english]{babel} \usepackage{amsmath}
\usepackage{amssymb} \usepackage{amsfonts} \usepackage{makecell}
\usepackage{hyperref} \usepackage[draft,deletedmarkup=xout]{changes}
\usepackage{mathtools} \usepackage{soul} \usepackage{ulem}
\usepackage{relsize}

\definecolor{green}{rgb}{0.19,0.64,0.54}
\definecolor{blue}{rgb}{0,0,1} \definecolor{reddish}{rgb}{0.65, 0.2,
  0.2} \definecolor{darkgreen}{rgb}{0.2,0.7,0.3}
\definecolor{darkblue}{rgb}{0.3,0.40,0.48}
\definecolor{gray}{rgb}{.8,.8,.8}

\newcommand{\scalar}[2]{\langle\kern.3ex #1 \kern.3ex|\kern.3ex#2\kern.3ex\rangle}

\hypersetup{,
  pdftitle={Time},
  pdfauthor={H. Bergeron, P Malkiewicz and P Peter},
  colorlinks=true, 
  linkcolor=darkblue, 
  citecolor=darkgreen, 
  filecolor=reddish, 
  linktocpage=true
  }

\newcommand{\dd}{\mathrm{d}}
\newcommand{\ex}{\mathrm{e}}
\newcommand{\GN}{G_\textsc{n}}
\newcommand{\Rez}{\Re\mathrm{e}}

\newcommand{\ud}{\mathrm{d}}

\newcommand{\be}{\begin{equation}}
\newcommand{\ee}{\end{equation}}
\newcommand{\ba}{\begin{eqnarray}}
\newcommand{\ea}{\end{eqnarray}}

\begin{document}

\title{Quantum entanglement and non-Gaussianity in the primordial Universe}

\author{Herv\'e Bergeron} \email{herve.bergeron@universite-paris-saclay.fr}

\affiliation{Institut des Sciences Mol\'{e}culaires d'Orsay (ISMO), UMR 8214 CNRS, Universit\'{e} Paris-Saclay, 91405 Orsay Cedex, France}

\author{Przemys{\l}aw Ma{\l}kiewicz}
\email{Przemyslaw.Malkiewicz@ncbj.gov.pl}

\affiliation{National Centre for Nuclear Research, Pasteura 7, 02-093
  Warszawa, Poland}

\author{Patrick Peter} \email{peter@iap.fr}

\affiliation{${\cal G}\mathbb{R}\varepsilon\mathbb{C}{\cal O}$ --
  Institut d'Astrophysique de Paris, CNRS and Sorbonne Universit\'e,
  UMR 7095 98 bis boulevard Arago, 75014 Paris, France}

\begin{abstract}
We propose a new method to investigate signatures of a quantum gravity phase in the primordial state of cosmological perturbations. We formulate and study a quantum model of a perturbed Friedmann-Lema\^{\i}tre-Robertson-Walker universe beyond a tensor-product Born-Oppenheimer-like factorization, that is, without restricting the wave function of the universe to the product of the background and perturbation wave functions. We show that the quantum dynamics generically does not preserve the product form of the universe's wave function, which spontaneously evolves into a more general entangled state. Upon expanding this state in a suitable basis of background wave functions and setting Gaussian initial conditions for the perturbations, we numerically find that each of these wave functions becomes associated with a non-Gaussian state of an inhomogeneous perturbation.
\end{abstract}

\maketitle

\section{Introduction}
The idea of quantum origin of the primordial structure in the Universe has been extensively studied for several decades. Its most prominent implementation, the theory of inflation~\cite{Guth:1980zm,Mukhanov:2005sc,Peter:2013avv,Martin:2024qnn}, assumes a classical background spacetime on which quantum fluctuations of the metric and matter fields propagate. However, this approach overlooks the quantum characteristics of the global geometry. It has been conjectured~\cite{Borde:2001nh} that inflationary spacetimes are past incomplete (see, however, Ref.~\cite{Lesnefsky:2022fen}). Consequently, various models have been developed that incorporate a quantum background geometry, potentially resolving the initial singularity and offering a dynamically complete depiction of the universe. While some of these models undergo an inflationary phase in their evolution, others serve as alternative solutions regarding the origin of the primordial Universe.

One common characteristic shared by all known quantum models is that they always rely on the Born-Oppenheimer-type dynamics (see, e.g., \cite{Peter_2006,Pinho_2007,PhysRevD.79.064030, Ma_kiewicz_2021, Kiefer:2004xyv, PhysRevD.87.104008, PhysRevD.103.066005, KAMENSHCHIK2013518, Gomar_2014,PhysRevD.105.086014}). Specifically, the wave function of the universe is constrained to be the product of the wave functions for the background geometry and for each perturbation mode, with every degree of freedom maintaining a definite state at all times. This is a highly restrictive assumption as a generic quantum gravity state should exhibit some amount of entanglement. Moreover, in quantum cosmology, the universe is not in a stationary state but undergoes rapid expansion, and possibly contraction, with potentially substantial interaction between the perturbation modes and the background.

The goal of this work is to make the first step toward a more complete description of the dynamics of quantum cosmological systems. Herein, we consider a quantum bounce model of a Friedmann-Lema\^{\i}tre-Robertson-Walker (FLRW) universe filled with a perfect fluid and furnished with scalar perturbations. We study the dynamics of this system, setting it initially on the contracting branch in a Born-Oppenheimer (BO) state. By BO state we mean a pure time-dependent tensor-product state: background state $\otimes$ perturbation state\footnote{Note our wording may be contrasted with the usual BO formulation as initially developed in Refs.~\cite{Born:1927rpw,Mott:1931}.} (see below). We follow the development of this quantum state as the universe contracts, bounces and reexpands. We examine the final entangled state and identify the potential observational signatures of the quantum dynamics beyond the Born-Oppenheimer-like factorization.

The physical interpretation that we apply to our formalism implies the type of corrections which we obtain. In spite of the wave function being a superposition of many background states, all our observables are limited to a single branch of a semiclassical FLRW universe evolving according to the background-level Hamiltonian. We do not consider the backreaction of the perturbation on this semiclassical state, assuming it is either negligible or already incorporated into the background dynamics. Rather, we focus on the effect of other backgrounds in the virtual branches of the wave function on the evolution of the perturbation itself which, as stated above, is measured only in the ``observable'' branch. We use a quantum bounce model for this investigation, so the obtained results may seem specific to the quantum bounce scenario. However, we believe that the discussed effects should be seen more broadly as coming from a quantum gravity phase of the universe and could result in a nonstandard initial condition for the inflationary models as well.

To facilitate the analysis, we restrict the Hilbert space as much as possible, allowing nevertheless for the entanglement effects to be present. We also make use of a recently discovered new class of coherent states on the half line to conveniently represent the dynamics of the background geometry states.

The plan of this work is as follows. In Sec.~\ref{model}, we introduce the classical model, quantize it and use the Born-Oppenheimer basis to formulate its complete quantum dynamics. Section~\ref{Biverse} is devoted to applying the framework to the case of two background states and their associated respective perturbation states. This simplified Hilbert space is established with the use of exact coherent states describing the background and represents the simplest setup in which the deviations from the BO dynamics can be studied. Section~\ref{results} presents the results of a numerical simulation while Section~\ref{Discussion} discusses our interpretation of the results and contains some concluding remarks.

\section{The model}\label{model}

\subsection{Classical Hamiltonian formalism}

In this work, we study the model of scalar perturbations in a flat FLRW universe filled with a single barotropic fluid. The derivation of the physical (reduced) Hamiltonian for this model can be found, e.g., in \cite{Mukhanov:1990me,Pinho:2006ym,Ma_kiewicz_2019,Martin_2022} where the physical dynamics is defined with respect to a fluid variable, denoted by $\tau$ and playing the role of internal time. For the present work, we find it more convenient to introduce another internal time, denoted by $\eta$, corresponding to the usual conformal time. We skip here the details of this simple but tedious redefinition and only provide the final definitions of internal time, basic variables, and physical Hamiltonian. 

The background dynamical variables $(q,p)\in\mathbb{R}^+\times\mathbb{R}$ read~\cite{Martin:2022ptk}
\begin{equation}\begin{split}
q&=\frac{4\sqrt{6}}{(1+3w)\sqrt{1+w}}a^{(1+3w)/2}
\equiv \gamma a^{(1+3w)/2},\\
p&=\frac{\sqrt{6(1+w)}}{2\kappa_0} a^{3(1+w)/2}H,\end{split}
\label{qpa}
\end{equation}
where $H=\dot{a}/(Na)$ is the Hubble rate, $\kappa_0$ is a rescaled gravitational constant\footnote{With $\kappa = 8\pi\GN$ and $\mathcal{V}_0$ the volume of the compact coordinate spatial volume, one has $\kappa_0=\kappa/\mathcal{V}_0$.} and $w$ is the barotropic index, i.e., the ratio of the pressure to the energy density of the fluid. We note that $p$ is a constant of motion.

In terms of the internal conformal time $\eta$, the dynamical variables $(q,p)$ form a canonical pair and the physical Hamiltonian reads
\begin{align}\label{new H}\begin{split}
H&=H^{(0)}+\sum_{\bm{k}} H^{(2)}_{\bm{k}},~~H^{(0)}= (2\kappa_0) {p}^2,\\
H^{(2)}_{\bm{k}}&=\frac12 |\pi_{v,{\bm{k}}}|^2+\frac12 \left( w
  k^2 -  \mathcal{V}\right) |v_{\bm{k}}|^2, \end{split}
\end{align}
where $$\mathcal{V}= \frac{8(2\kappa_0)^2 (1-3w)}{(1+3w)^2}\frac{{p}^2}{{q}^2}$$ and the perturbation variable $v_{\bm{k}}$ is the usual Mukhanov-Sasaki variable. One can easily check that the above Hamiltonian generates the mode dynamics
\begin{equation}\label{eomP}
v''_{\bm{k}} + \left( w k^2 - \mathcal{V}
\right) v_{\bm{k}} = 0
\end{equation}
for $v_{\bm{k}}$, where a prime denotes a derivative with respect to $\eta$ and $\mathcal{V}=a''/a$.

\subsection{Quantization}
The usual canonical prescription involves the replacements:
$$q\rightarrow\widehat{Q},~~p\rightarrow\widehat{P},$$
which in the case of compound dynamical variables $f(q,p)$ is followed by symmetrization with respect to $\widehat{Q}$ and $\widehat{P}$. However, it turns out that the momentum operator $\widehat{P}$ defined on the half line is not a self-adjoint operator. For this reason, we shall apply an enhanced prescription \cite{Klauder:2015ifa} that involves the replacements:
$$q\rightarrow\widehat{Q},~~d=qp\rightarrow\widehat{D}=\frac{1}{2}(\widehat{Q}\widehat{P}+\widehat{P}\widehat{Q}),$$
which in the case of compound dynamical variables $f(q,q^{-1}d)$ is followed by the symmetrization with respect to $\widehat{Q}$ and $\widehat{D}$. Both the position $\widehat{Q}$ and dilation $\widehat{D}$ operators are self-adjoint and they generate a unitary representation of the group of affine transformations on the real line \cite{Gazeau:2015nkc}.

Quantization of the background Hamiltonian $H^{(0)}$ with the enhanced prescription yields
\begin{align}\label{bgham}
H^{(0)}= (2\kappa_0) \left(\frac{d}{q}\right)^2\rightarrow ~\widehat{H}^{(0)}=(2\kappa_0)\left(\widehat{P}^2+\frac{K}{\widehat{Q}^2}\right),
\end{align}
where the constant $K>0$ depends on the operator ordering~--~see Eq. (76) of \cite{Ma_kiewicz_2020} for more details~--~; we shall assume $K>3/4$ to guarantee a unique self-adjoint realization of the above Hamiltonian. In what follows, we set $\kappa_0=\frac{1}{2}$.

The gravitational potential $\mathcal{V}$ of the perturbation Hamiltonian $H^{(2)}$ is then quantized through the replacement
\begin{align}
\mathcal{V}\propto\frac{{d}^2}{{q}^4}~\rightarrow ~\widehat{\mathcal{V}}\propto\widehat{D}\widehat{Q}^{-4}\widehat{D},
\end{align}
where a computationally convenient operator ordering has been chosen. We have checked that there exists a unique self-adjoint realization of the above quantum potential.

\subsection{Background dynamics}\label{bd}

For describing quantum states of the background, we find it useful to use a coherent state representation rather than the usual one based on self-adjoint operators such as the position (i.e., the scale factor) or the energy. We shall denote the coherent states by $|q,p\rangle$, defined
with the help of a fiducial state $|\xi\rangle$ 
through~\cite{Bergeron:2023zzo}
\begin{equation}
|q,p\rangle = \ex^{-i\phi} \exp\left( \frac{i p}{2q}\hat{Q}^2 \right)
\exp\left(- i \ln q \hat{D}\right) |\xi\rangle,
\label{qpstate}
\end{equation}
where
$$
\ex^{-i\phi} = \left( \frac{\xi_{\nu} - i q p}{\xi_{\nu} + i q p} \right)^{\frac{\nu+1}{2}},
$$
and $\nu>\frac{1}{2}$ and $\xi_{\nu}=\left[ \Gamma(\nu+\frac{3}{2})/\Gamma(\nu+1)\right]^2$. Such states resolve the unity 
\begin{align}
    \int |q,p\rangle\langle q,p|~\frac{\ud q\ud p}{2\pi c_0} =\mathbf{1},
\end{align}
with $c_0$ a normalization factor built as an integral over
the fiducial wave function $\xi(x) = \langle x|\xi\rangle$, explicitly
$$
c_0 = \int_0^{+\infty} \frac{\dd x}{x^2} |\xi(x)|^2.
$$
Such states permit one to express any quantum state as a wave function of $(q,p)$, namely
\begin{align}
    |\Psi\rangle \mapsto \Psi(q,p)=\frac{\langle q,p |\Psi\rangle}{\sqrt{2\pi c_0}},\label{norm}
\end{align}
with the normalization condition $\int |\Psi(q,p)|^2\ud q\ud p=1$. Consequently, the quantity
\begin{align}
    \rho_{\Psi}(q,p)=|\Psi(q,p)|^2
\end{align}
is the phase space probability density for the state $\Psi$. For simplicity, in what follows, we assume $2\pi c_0 \to 1$.

The coherent states themselves can serve as a kind of semiclassical states for the background geometry. The phase space probability density for a quantum state $|\tilde{\Psi}\rangle=|\tilde{q},\tilde{p}\rangle$ simply reads
\begin{align}
    \rho_{\tilde{q},\tilde{p}}(q,p)=|\langle q,p |\tilde{q},\tilde{p}\rangle|^2.
\end{align}
This formula emphasizes the fact that coherent states are not orthogonal to one another and thus form an overcomplete basis in the Hilbert space. In this work we shall use the coherent states defined on the half line $x>0$ as
\begin{widetext}
\begin{align}\label{ACS}
\scalar{x}{q,p} = \sqrt{\frac{2}{\Gamma(\nu+1)}} \left( \frac{\xi_{\nu} - i q p}{\xi_{\nu} + i q p} \right)^{\frac{\nu+1}{2}}  \xi_{\nu}^{\frac{\nu+1}{2}} \, \frac{x^{\nu+1/2}}{q^{\nu+1}} \exp \left[ - \frac{1}{2} (\xi_{\nu} - i \, q p) \frac{x^2}{q^2} \right].
\end{align}
\end{widetext}
It has recently been shown \cite{Bergeron:2023zzo} that within the set
of coherent states defined above, there exist one-parameter families that solve the Schr\"odinger equation involving the Hamiltonian \eqref{bgham}, namely
\begin{align}
i\frac{\partial}{\partial\eta}|q(\eta),p(\eta)\rangle=\widehat{H}^{(0)}|q(\eta),p(\eta)\rangle,
\label{Sch0}
\end{align}
where $\nu=\sqrt{K+\frac{1}{4}}$, while $q(\eta)$ and $p(\eta)$ solve the Hamilton equations derived from the semiclassical Hamiltonian
\begin{align}
{H}^{(0)}_\text{sem}=\frac{\xi_{\nu}}{\nu+1}\langle q,p |\widehat{H}^{(0)}|q,p\rangle=p^2+\frac{\xi_{\nu}^2}{q^2}.
\end{align} 
Thus, the evolution of the variables $q(\eta)$ and $p(\eta)$ reads
\begin{align}\label{semi}\begin{split}
q(\eta)&=q_\textsc{b}\sqrt{1+\left( \frac{\eta-\eta_\textsc{b}}{\bar{\eta}}\right)^2},\\
p(\eta)&=\frac12 \frac{\dd q}{\dd\eta} = \frac{q_\textsc{b}^2}{2 \bar{\eta} q(\eta)}
\left( \frac{\eta-\eta_\textsc{b}}{\bar{\eta}}\right),\end{split}
\end{align}
representing solutions for the scale factor according to Eq.~\eqref{qpa}. 
These solutions exhibit a bouncing behavior at an arbitrary time $\eta_\textsc{b}$ at which the scale factor reaches its minimal
value $a_\textsc{b} = (q_\textsc{b}/\gamma)^{2/(1+3w)}$. Denoting
by $E$ the energy $E=p^2(\eta)+\xi_{\nu}^2/q^2(\eta)$, constant
along such a semiclassical trajectory, one obtains the other 
relevant parameters, namely the minimum scale as
$q_\textsc{b}=\xi_{\nu}/\sqrt{E}$ and 
the typical bounce duration as $\bar{\eta}=\xi_{\nu}/E$.

\subsection{Complete dynamics}

To determine the dynamics of the full state including
perturbations on top of the background, one needs to solve the Schr\"odinger equation
\begin{align}\label{fullshr}
i\frac{\partial}{\partial\eta}|\Psi\rangle=\left(\widehat{H}^{(0)}+\widehat{H}^{(2)}\right)|\Psi\rangle,
\end{align}
where $\widehat{H}^{(2)}=\sum_{\bm{k}} \widehat{H}^{(2)}_{\bm{k}}$. The standard approach to solving this equation relies on the so-called Born-Oppenheimer approximation, which assumes the wave function to be the product of the background and the perturbation wave functions, namely 
$\Psi(q,\{ v_{\bm{k}} \}) = \psi^{(\text{bg})} (q) \psi^{(\text{pert})}(q,\{ v_{\bm{k}} \})$ and in which one introduces a small expansion parameter $\epsilon$, usually a mass ratio, thanks to which the equations can be solved iteratively order by order in powers of $\epsilon$; see, e.g., Ref.~\cite{Kiefer:2004xyv}, Sec. 5.4, for details. Here, we make use of a slightly different definition of a BO state based on a tensor product: we assume that a BO state reads
\begin{equation}
|\Psi^{\textsmaller{(}\textsc{bo}\textsmaller{)}}\rangle=|\psi^{(\text{bg})}(q) \rangle \otimes |\psi^{(\text{pert})} \left[\{ v_{\bm{k}} \};q\left(\eta\right),p\left(\eta\right) \right]\rangle,
\end{equation}
in which the background state $|\psi^{(\text{bg})}(q) \rangle $ depends on the background dynamical variable $q$ and satisfies the background Schr\"odinger equation \eqref{Sch0}, providing the semiclassical solution \eqref{semi} for $q(\eta)$ and $p(\eta)$. This solution is then plugged into the perturbation wave function as a time-varying parameter, and one finally gets the effective perturbation Schr\"odinger equation
\begin{align}\label{BOdynamics}
i\frac{\partial}{\partial\eta}|\psi^{(\text{pert})}
\rangle=\langle\psi^{(\text{bg})} |\widehat{H}^{(2)}
|\psi^{(\text{bg})}\rangle \, |\psi^{(\text{pert})}\rangle,
\end{align}
where the quantum gravitational potential $\widehat{\mathcal{V}}$ is replaced by the quantum average $\langle\psi^{(\text{bg})} |\widehat{\mathcal{V}}|\psi^{(\text{bg})}\rangle$ or another $c$-number that includes a quantum correction. Note that given our assumption of the tensor product, Eq.~\eqref{BOdynamics} means that the perturbation part of a BO state only evolves through the diagonal part  $\langle\psi^{(\text{bg})}|\widehat{H}^{(2)}
|\psi^{(\text{bg})}\rangle$ of the operator $\widehat{H}^{(2)}$. Therefore, the effect of the off-diagonal terms such as $\langle\psi_1^{(\text{bg})}|\widehat{H}^{(2)}|\psi_2^{(\text{bg})}\rangle$ is neglected, and this corresponds to our definition of the ``Born-Oppenheimer approximation''. Any exact solution of \eqref{fullshr} must take into account the off-diagonal terms $\langle\psi_1^{(\text{bg})}|\widehat{H}^{(2)}|\psi_2^{(\text{bg})}\rangle$ and can be written as a time-dependent linear superposition of different BO states.

The standard approach thus neglects the entanglement between different background and perturbation wave functions. However, starting with a simple product state, the entanglement must eventually be produced by gravitational dynamics, which couples perturbations to the background. Therefore, starting as
$$|\Psi\rangle=|\psi^{(\text{bg})}_0\rangle |\psi^{(\text{pert})}_0\rangle,$$
the quantum state after some time is expected to read
\begin{align}\label{generalstates}
|\Psi(\eta)\rangle=\sum_n |\psi_n^{(\text{bg})}(\eta)\rangle |\psi_n^{(\text{pert})}(\eta)\rangle,
\end{align}
with a nonvanishing state vector $|\psi_n^{(\text{bg})}(\eta)\rangle |\psi_n^{(\text{pert})}(\eta)\rangle$ for some $n>0$ \cite{Ma_kiewicz_2021}. The background states $|\psi_n^{(\text{bg})}(\eta)\rangle$ ($n=0,1,\dots$) can be assumed to solve the background Schr\"odinger equation and form a (time-dependent) basis in the background Hilbert space. The dynamics of the perturbation states $|\psi_n^{(\text{pert})}(\eta)\rangle$ is nontrivial and can in principle depend on other perturbation states $|\psi_m^{(\text{pert})}(\eta)\rangle$ with $m\neq n$. 

Upon substituting the general state \eqref{generalstates} into the full Schr\"odinger equation \eqref{fullshr} we find
\begin{align}\label{generaldynamics}
i\frac{\partial}{\partial\eta}|\psi_n^{(\text{pert})}\rangle=
\sum_{\ell m} M_{n\ell}^{-1}\widehat{H}^{(2)}_{\ell m} |\psi_m^{(\text{pert})}\rangle,
\end{align}
where $\widehat{H}^{(2)}_{\ell m}=\langle\psi_\ell^{(\text{bg})} |\widehat{H}^{(2)}|\psi_m^{(\text{bg})}\rangle$ is a matrix representation of $\widehat{H}^{(2)}$ in the background Hilbert space  and $M_{nm}=\langle\psi_n^{(\text{bg})} |\psi_m^{(\text{bg})}\rangle$ is the matrix of scalar products between the background solutions. We clearly see that assuming initially $|\psi_0^{(\text{pert})}(\eta)\rangle\neq 0$, the dynamics generically produces nontrivial $|\psi_n^{(\text{pert})}(\eta)\rangle$ for $n>0$ as long as $\langle\psi_n^{(\text{bg})} |\widehat{H}^{(2)}|\psi_0^{(\text{bg})}\rangle\neq 0$. Hence, the background universe starting out from a simple `semiclassical' state $|\psi_0^{(\text{bg})}(\eta)\rangle$ eventually evolves into a multibranched state \eqref{generalstates}. 

We conclude that a quantum multiverse naturally emerges in this approach even though no ``third quantization'' \cite{Caderni:1984pw} is performed on the gravitational field. However natural the presence of the multiverse may appear from the point of view of ordinary quantum mechanics, it poses a severe interpretational problem in quantum cosmology and complicates taking the classical limit in which the observable universe must be recovered. The viewpoint which we adopt in this work is that the fixed background solution $|\psi_0^{(\text{bg})}(\eta)\rangle$ and its associated perturbation state $|\psi_0^{(\text{pert})}(\eta)\rangle$ underlie our classical universe with its inhomogeneous perturbation in the sense that at late times the latter provides a reasonable approximation to (or, realization of) the quantum solution $|\psi_0^{(\text{bg})}(\eta)\rangle |\psi_0^{(\text{pert})}(\eta)\rangle$. All other background solutions $|\psi_n^{(\text{bg})}(\eta)\rangle$ for $n>0$ are ``virtual'' with their only role to generate an interaction loop $|\psi_0^{(\text{pert})}(\eta)\rangle\leftrightarrow |\psi_n^{(\text{pert})}(\eta)\rangle$ in the dynamics of the ``observed'' perturbation $|\psi_0^{(\text{pert})}(\eta)\rangle$ belonging to the physical background universe $|\psi_0^{(\text{bg})}(\eta)\rangle$, according to the dynamical law \eqref{generaldynamics}.

\begin{figure}[t!]
    \centering
   \includegraphics[width=0.22\textwidth]{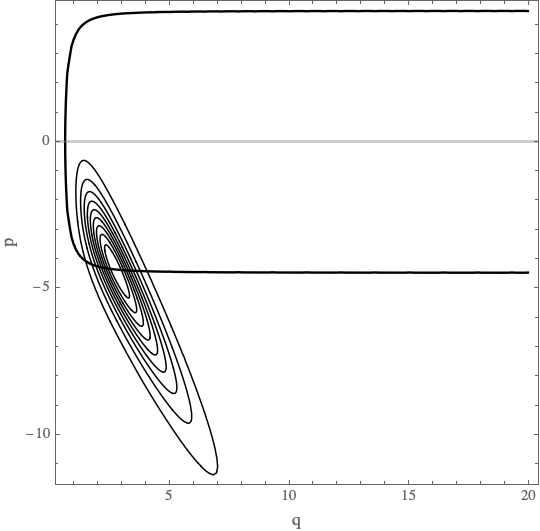}
    \includegraphics[width=0.22\textwidth]{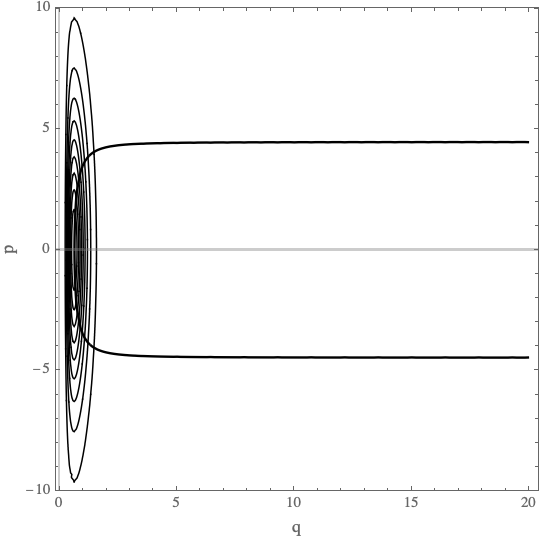}
    \includegraphics[width=0.22\textwidth]{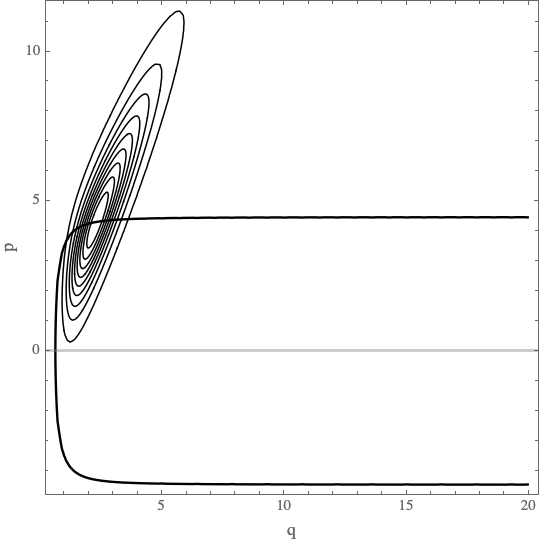}
    \includegraphics[width=0.22\textwidth]{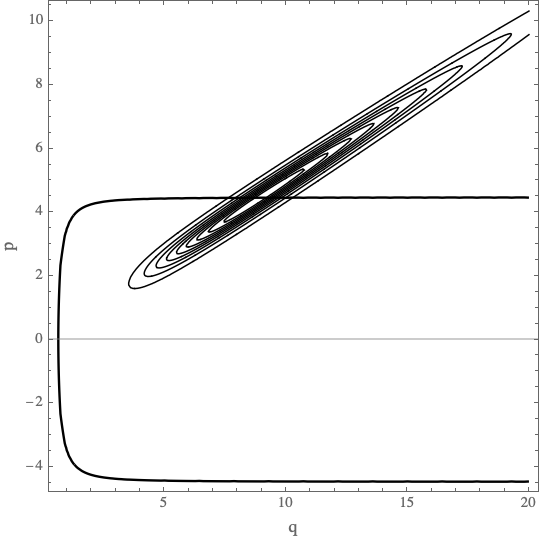}
    \caption{The bouncing evolution of the probability distribution (from top left to bottom right) of a coherent state \eqref{ACS} for $\nu=2$, $E=20$ and $\eta=-0.3,0,0.25,1$.}\label{fig1}
\end{figure}

\begin{figure}[t!]
    \centering
   \includegraphics[width=0.22\textwidth]{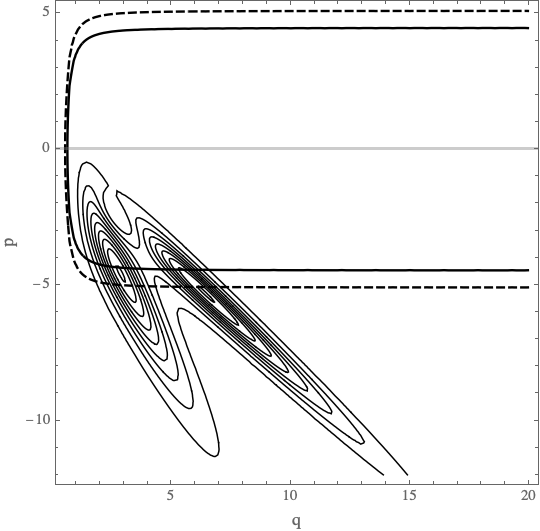}
    \includegraphics[width=0.22\textwidth]{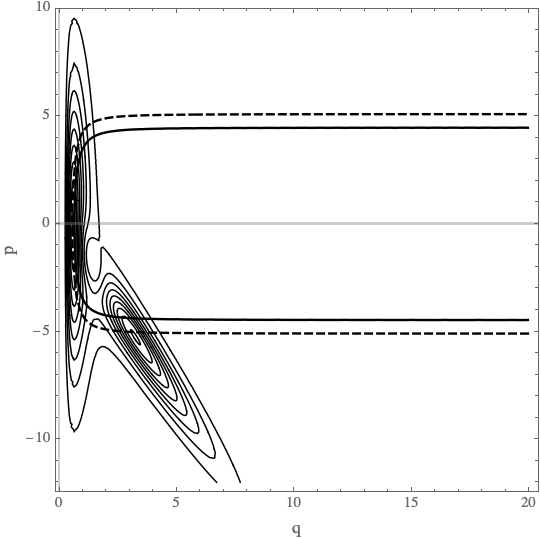}
    \includegraphics[width=0.22\textwidth]{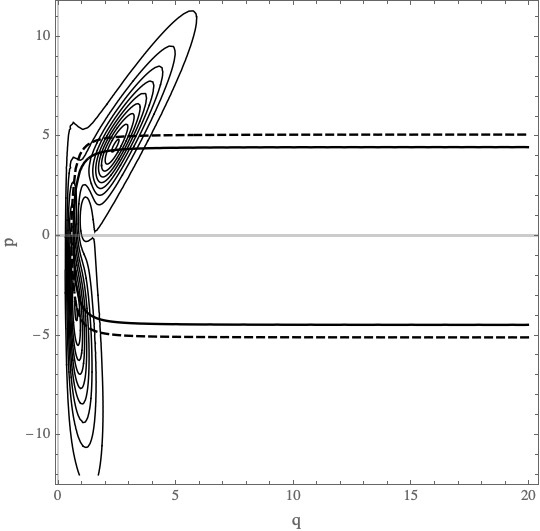}
    \includegraphics[width=0.22\textwidth]{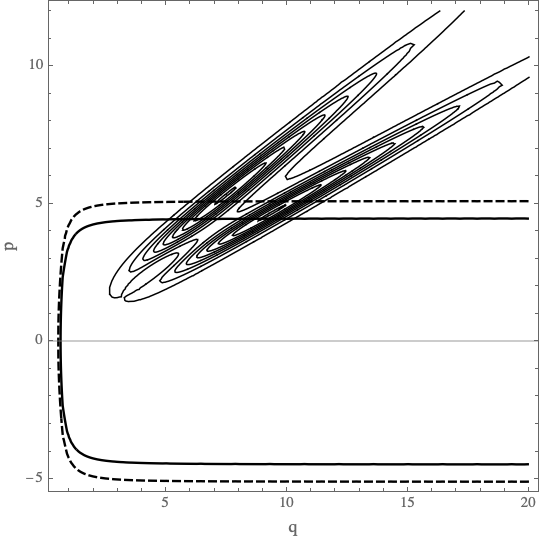}
    \caption{The evolution of the probability distribution (from top left to bottom right) of a combination of two coherent states \eqref{ACS} with the energies and the bouncing times respectively  $E=20,26$ and $\eta_b=0,0.3$, for $\nu=2$ and $\eta=-0.3,0,0.25,1$.}\label{fig3}
\end{figure}

Before concluding this section let us reformulate the general dynamics \eqref{generaldynamics} in terms of a fixed basis $\{|\phi_m\rangle\}$. Because the perturbation space of states is a Hilbert space, we can safely assume that there exists such a basis. We further demand that each of the basis element evolves according to the Born-Oppenheimer dynamical prescription with the background solution sharing the same label $n$, that is,
\begin{align}
i\partial_{\eta}|\phi_n\rangle=\langle\psi_n^{(\text{bg})} |\widehat{H}^{(2)}|\psi_n^{(\text{bg})} \rangle |\phi_n\rangle.
\label{dphin}
\end{align}
so that the physical perturbation state can be decomposed into
 \begin{align}
|\psi_n^{(\text{pert})}\rangle= \sum_m \alpha_{nm}|\phi_m\rangle,
\label{phim}
 \end{align}
for all $n$, with $\alpha_{nm}$ a set of time-dependent complex parameters to be determined below.

The entire dynamics is now encoded in the time development of $\alpha_{nm}(\eta)$, and one finds that they satisfy the dynamical law
\begin{align}\label{generaldynamics2}\begin{split}
i{\alpha}_{nm}'=\sum_{\ell r k p} M_{n\ell}^{-1}S_{mk}^{-1}{H}^{(2)}_{\ell r kp}\alpha_{rp}-\sum_{k p} S_{mk}^{-1}{H}^{(2)}_{ppkp}\alpha_{np},
\end{split}
\end{align}
where ${H}^{(2)}_{\ell mkp}=\langle\phi_k |\widehat{H}^{(2)}_{\ell m}|\phi_p \rangle$ is a complete matrix representation of the operator $\widehat{H}^{(2)}$ and $S_{kp}=\langle\phi_k|\phi_p \rangle$ is the matrix of scalar products between the perturbation solutions. The matrices $\widehat{H}^{(2)}_{\ell m}$ and $M_{n\ell}$ are defined below Eq. \eqref{generaldynamics}. 

Our goal in the next section is to study a specific example and verify if condition $\langle\psi_0^{(\text{bg})} |\widehat{H}^{(2)}|\psi_n^{(\text{bg})}\rangle\neq 0$ for $n>0$ can indeed hold and lead to a substantial modification of the Born-Oppenheimer dynamics \eqref{BOdynamics}.

\section{Biverse}\label{Biverse}

In what follows, we study only a single mode of perturbation $\bm{k}$, and restrict the cosmological setup to two independent background solutions $|\psi_0^{(\text{bg})}(\eta)\rangle=|q_0(\eta),p_0(\eta)\rangle$ and $|\psi_1^{(\text{bg})}(\eta)\rangle=|q_1(\eta),p_1(\eta)\rangle$, with respective energies $E_0$ and $E_1$. We use the respective perturbation solutions to the Born-Oppenheimer dynamics \eqref{BOdynamics}, denoted here by  $|\phi_{E_0}(\eta)\rangle$ and $|\phi_{E_1}(\eta)\rangle$, to form a two-dimensional basis for the perturbation states, namely $|\psi_0^{(\text{pert})}(\eta)\rangle=\alpha_{00}(\eta)|\phi_{E_0}(\eta)\rangle+\alpha_{01}(\eta)|\phi_{E_1}(\eta)\rangle$ and $|\psi_1^{(\text{pert})}(\eta)\rangle=\alpha_{10}(\eta)|\phi_{E_0}(\eta)\rangle+\alpha_{11}(\eta)|\phi_{E_1}(\eta)\rangle$,
where  $|\phi_{E_0}(\eta)\rangle$ and $|\phi_{E_1}(\eta)\rangle$ are fixed by 
Eq.~\eqref{dphin}, which reads in this case
\begin{align}\label{BOdynamics2}\begin{split}
i\partial_{\eta}|\phi_{E_0}\rangle&=\langle q_0(\eta),p_0(\eta)|\widehat{H}^{(2)}_{\bm{k}}|q_0(\eta),p_0(\eta)\rangle |\phi_{E_0}\rangle,\\
i\partial_{\eta}|\phi_{E_1}\rangle&=\langle q_1(\eta),p_1(\eta)|\widehat{H}^{(2)}_{\bm{k}}|q_1(\eta),p_1(\eta)\rangle |\phi_{E_1}\rangle,\end{split}
\end{align}
with the asymptotic vacuum state set as the initial state. The complete state thus reads
\begin{align}\label{bistate}\begin{split}
|\Psi(\eta)\rangle&=|q_0,p_0\rangle\left[\alpha_{00}(\eta)|\phi_{E_0}(\eta)\rangle+\alpha_{01}(\eta)|\phi_{E_1}(\eta)\rangle\right]\\
&+|q_1,p_1\rangle\left[\alpha_{10}(\eta)|\phi_{E_0}(\eta)\rangle+\alpha_{11}(\eta)|\phi_{E_1}(\eta)\rangle\right].\end{split}
\end{align}
The dynamical law \eqref{generaldynamics2} is then applied to the state \eqref{bistate} to derive the dynamics for the amplitudes $\alpha_{00}$, $\alpha_{01}$, $\alpha_{10}$ and $\alpha_{11}$. 

Working in the Schr\"odinger representation with $H^{(2)}_{\bm{k}}$ defined in \eqref{new H}, we find, with $i,j\in \{0,1\}$,
\begin{align}\begin{split}
\hat{H}_{ij}^{(2)}=\left(-\frac{1}{2}\frac{\partial^2}{\partial v_{\bm{k}}^2} + \frac{1}{2}wk^2v_{\bm{k}}^2\right)M_{ij}-\frac{1}{2}\mathcal{V}_{ij} v_{\bm{k}}^2,
\end{split}\end{align}
where $\mathcal{V}_{ij}=\langle q_i,p_i |\widehat{\mathcal{V}}| q_j,p_j\rangle$ and
$$
M_{ij} := \langle q_i,p_i  | q_j,p_j \rangle = \int \dd x \langle q_i,p_i | x \rangle \langle x |q_j,p_j \rangle,
$$
whose integration leads, through the explicit form \eqref{ACS}, to
\begin{align}
M_{ij}=\left[\frac{2\left(\displaystyle\frac{\xi_{\nu}+ip_iq_i}{\xi_{\nu}-ip_iq_i}\right)^{1/2}\left(\displaystyle\frac{\xi_{\nu}-ip_jq_j}{\xi_{\nu}+ip_jq_j}\right)^{1/2}}{\left(\displaystyle\frac{q_j}{q_i}+\frac{q_i}{q_j}\right)
+\displaystyle\frac{i}{\xi_{\nu}}\left( p_iq_j-p_jq_i\right)}\right]^{\nu+1}.
\label{Mij}
\end{align}
It is also possible to obtain the 
potential $\mathcal{V}_{ij}$ analytically. Using 
Eq.~\eqref{qpstate} and the commutation relation $[\hat{Q},\hat{D}]=i\hat{Q}$, one finds
$$
\hat{D}e^{i\phi}|q,p\rangle = i\left(q\frac{\partial}{\partial q}-p\frac{\partial}{\partial p} \right)e^{i\phi} |q,p\rangle,$$
\noindent which permits one to calculate $\hat{D}|q,p\rangle$ and subsequently
the necessary matrix elements for the potential $\langle q_i,p_i|\hat{D}\hat{Q}^{-4}\hat{D}| q_j,p_j \rangle$ through
\begin{widetext}
$$\langle q_i,p_i|\hat{D}\hat{Q}^{-4}\hat{D}| q_j,p_j \rangle=e^{i(\phi_i-\phi_j)}\left(q_i\frac{\partial}{\partial q_i}-p_i\frac{\partial}{\partial p_i} \right)\left(q_j\frac{\partial}{\partial q_j}-p_j\frac{\partial}{\partial p_j} \right)e^{i(\phi_j-\phi_i)}\langle q_i,p_i|\hat{Q}^{-4}| q_j,p_j \rangle .$$
Using the representation \eqref{ACS}, this leads to the Hermitian matrix
\begin{align}\begin{split}
\mathcal{V}_{ij}=\frac{2 (1-3w)}{(1+3w)^2 \nu(\nu-1)}\left\{(\nu+1)(\nu-3)\left[\frac{p_i^2}{q_i^2}+\frac{p_j^2}{q_j^2}+2i\xi_{\nu}\left(\frac{p_i}{q_i^3}-\frac{p_j}{q_j^3}\right)-\xi_{\nu}^2\left(\frac{1}{q_i^4}+\frac{1}{q_j^4}\right)\right]\right.\\
\left.+(\nu^2+3)\left[2\frac{p_i}{q_i}\frac{p_j}{q_j}-2i\xi_{\nu}\left(\frac{p_i}{q_iq_j^2}-\frac{p_j}{q_jq_i^2}\right)+\frac{\xi_{\nu}^2}{q_i^2 q_i^2}\right]\right\} M_{ij},\end{split}
\label{Vij}
\end{align}
which, for $i=j$, reduces to
\begin{align}\label{potii}
\mathcal{V}_{ii}=\frac{8 (1-3w)}{(1+3w)^2}\left[ \frac{p_i^2}{q^2_i}+\frac{\nu+3}{\nu(\nu-1)}\frac{\xi_{\nu}^2}{q_i^4}\right].
\end{align}
Furthermore, assuming the following Gaussian perturbation
representation
\begin{equation}
\langle v_{\bm{k}}|\phi_{E_i}\rangle= \left[ \frac{2\Rez(\Omega_i)}{\pi} \right]^{1/4}
\exp\left[ -\Omega_i(\eta) v_{\bm{k}}^2 \right],
\label{wave function}
\end{equation}
it is straightforward to calculate the matrix elements ${H}_{ijkl}^{(2)}$. One obtains
\begin{align}\begin{split}
H_{ijkl}^{(2)}=\sqrt{2}M_{ij}\left[ \Rez\left(\Omega_k\right)
\Rez\left(\Omega_l\right)\right]^{\frac14}\left[\frac{\Omega_l}{(\Omega^\star_k+ \Omega_l)^{\frac12}}-\frac{ \Omega_l^2}{(\Omega^\star_k+ \Omega_l)^{\frac32}}+\frac{ w k^2-\mathcal{V}_{ij}/M_{ij}}{4(\Omega^\star_k+ \Omega_l)^{\frac32}}\right].
\end{split}\end{align}
\end{widetext}
The Born-Oppenheimer equations \eqref{BOdynamics2} can be rewritten in terms of the mode functions
$f_i(\eta)$ that satisfy the mode equation
\begin{align}\begin{split}
f''_i+(w k^2-\mathcal{V}_{ii})f_i=0.\end{split}
\end{align}
This mode function permits us to recover the variance through
\begin{align}
\Omega_i=-\frac{i}{2}\frac{f'_i}{f_i},
\end{align}
and therefore to reconstruct the perturbation wave function \eqref{wave function}.

\begin{figure}[t!]
    \centering
   \includegraphics[width=0.42\textwidth]{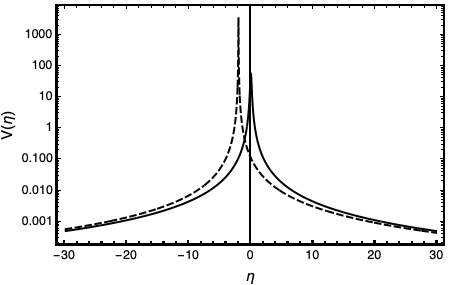}
 \caption{The evolution of the potentials $V_{00}(\eta)$ (full line) and $V_{11}(\eta)$ (dashed line) appearing in the mode equations \eqref{modes}. For illustrative purposes, in this figure and the following ones, we assume $w\to \frac16$ for definiteness. The potentials are seen to act mostly at the bounce location for both bounces.}\label{pots}
\end{figure}

\begin{figure}[t!]
    \centering
   \includegraphics[width=0.42\textwidth]{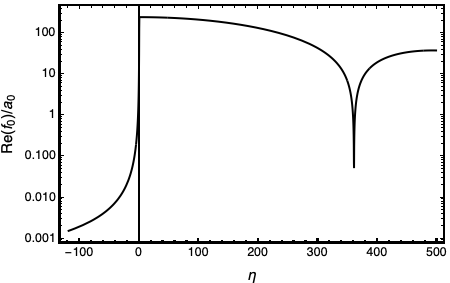}\vspace{0.2cm}\\
   \includegraphics[width=0.42\textwidth]{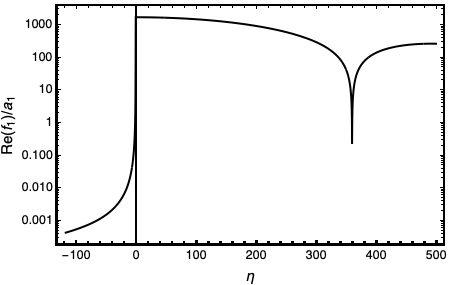}
 \caption{The evolution of $\Rez(f_0)/a_0$ and $\Rez(f_1)/a_1$ for $k=0.01$. The first perturbation gets amplified around the first background bounce at $\eta=0$, whereas the second one around the second background bounce at $\eta=-2$. The second background is more massive than the first one ($E_1>E_0$) and thus the second amplitude gets more amplified than the first, i.e., $\Rez(f_1)/a_1>\Rez(f_0)/a_0$.}\label{fig4}
\end{figure}

\section{Results}
\label{results}

The coherent states \eqref{ACS} represent a bouncing universe. In what follows, we use the family of time-independent coherent states $(q,p)\mapsto |q,p\rangle$ to give a probabilistic interpretation to the quantum state of the universe and its evolution. As can be seen from Fig. \ref{fig1}, the probability distribution (for a fixed $E$) that is obtained reads
\begin{align}
\rho_E(q,p,\eta)= |\psi_{E}(q,p,\eta)|^2 = 
\left|\int_0^{\infty} \dd x \langle q,p|x\rangle \langle x|\psi_{E} \rangle\right|^2,
\label{rhoE}
\end{align}
with $\psi_{E}(q,p,\eta)$ defined as in Eq.~\eqref{norm}, namely its ``$x$'' representation is given by Eq.~\eqref{ACS} with $q\mapsto q(\eta)$ and $p\mapsto p(\eta)$ from Eq.~\eqref{semi} corresponding to the value $E$ for the energy.

The density \eqref{rhoE} first approaches the singular boundary $q=0$ with $p<0$ (contraction), then bounces against the boundary (big bounce) and finally moves away from the boundary with $p>0$ (expansion). The distribution exhibits a simple shape and a simple dynamics that is symmetric with respect to the bounce. The background trajectory is obtained from the expectation values of $\widehat{Q}$ and $\widehat{P}$, denoted, respectively, as $q$ and $p$. As pointed out in Sec~\ref{bd}, the background trajectories solve the dynamics implied by the semiclassical Hamiltonian and are given by Eq. \eqref{semi}.

Suppose that the background wave function spreads into a combination of two coherent states with different energies and bouncing times. Then the probability distribution for such a ``biverse'' is a more complicated function of $q$, $p$ and $\eta$. In Fig. \ref{fig3} we plot a probability distribution for a biverse with equal weights for each component universe:
\begin{equation}
\rho_{E_0,E_1}(q,p,\eta)=\frac{|\psi_{E_0}(q,p,\eta-\eta_{0}) +
\psi_{E_1}(q,p,\eta-\eta_{1})|^2}{2(1+\Rez\langle q_0,p_0|q_1,p_1\rangle)}.
\end{equation}
The dynamics is no longer symmetric with respect to the bounce that happens at different times for different components. Interference between these two components is noticeable. Each component follows its semiclassical trajectory.

We now turn to the main numerical computation of the paper: a case of two Born-Oppenheimer universes. In the following, we assume the numerical values 
$\nu=2$, $E_0=10$, $E_1=80$, $\eta_{0}=0$, $\eta_{1}=-2$, and $k=0.01$.
We determine the perturbation wave function on each component universe $\langle v_{\bm{k}}|q_0,p_0 \rangle$ and $\langle v_{\bm{k}}|q_1,p_1\rangle$ by solving the mode equations for $f_0(\eta)$ and $f_1(\eta)$, namely
\begin{align}\label{modes}\begin{split}
f''_0+\left( wk^2- \langle q_0,p_0 | \widehat{\mathcal{V}}| q_0,p_0 \rangle \right) f_0=0,\\
f''_1+\left( wk^2-\langle q_1,p_1 |\widehat{\mathcal{V}}| q_1,p_1\rangle \right) f_1=0,\end{split}
\end{align}
with the initial condition corresponding to the asymptotic vacuum state: $f_i(-\infty)=(2\sqrt{w}k)^{-1/2}$ and $f'_i(-\infty)=i\sqrt{\sqrt{w}k/2}$. 
The potentials $V_{00}(\eta)=\langle q_0,p_0 | \widehat{\mathcal{V}}| q_0,p_0 \rangle $ and $V_{11}(\eta)=\langle q_1,p_1 |\widehat{\mathcal{V}}| q_1,p_1\rangle$ are plotted in Fig. \ref{pots}.

The behavior of the mode functions $f_0$ and $f_1$ for $k=0.01$ is plotted in Fig. \ref{fig4}. The maximum amplification of the amplitudes $\Rez(f_0)$ and $\Rez(f_1)$ occurs, respectively, at $\eta\approx -2$ and $\eta\approx 0$, which is consistent with the corresponding bounce times. In the second, more massive, background the amplitude amplification is stronger. These solutions form a basis on which we shall express the perturbation wave function \eqref{bistate} determined by the fuller dynamical law \eqref{generaldynamics2}.

\begin{figure}[t]
    \centering
   \includegraphics[width=0.42\textwidth]{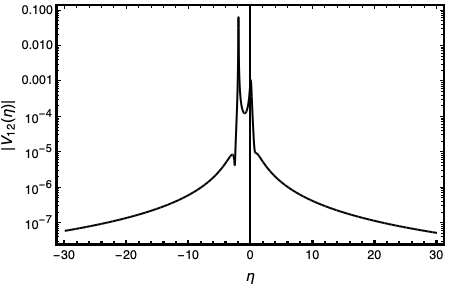}
 \caption{The evolution of the overlap $\langle \psi_{E_0} |\hat{\mathcal{V}}| \psi_{E_1}\rangle$ between the two component universes on the gravitational potential. This quantity is responsible for the interaction between the universes with their primordial structures.}\label{pot12}
\end{figure}

The background wave functions $| q_0,p_0 \rangle$ and $| q_1,p_1 \rangle$ are not orthogonal as their overlap (which is constant in time) is of order of $10^{-4}$. This fact is incorporated in the dynamics \eqref{generaldynamics} by the matrix element $M_{01}$. However, the key quantity in the process of entangling the two component universes is their overlap on the potential $V_{01}(\eta)=\langle q_0,p_0  |\hat{\mathcal{V}}| q_1,p_1 \rangle$ that is plotted in Fig. \ref{pot12}. Its value is largest at the bounces.

\begin{figure}[t]
    \centering
   \includegraphics[width=0.5\textwidth]{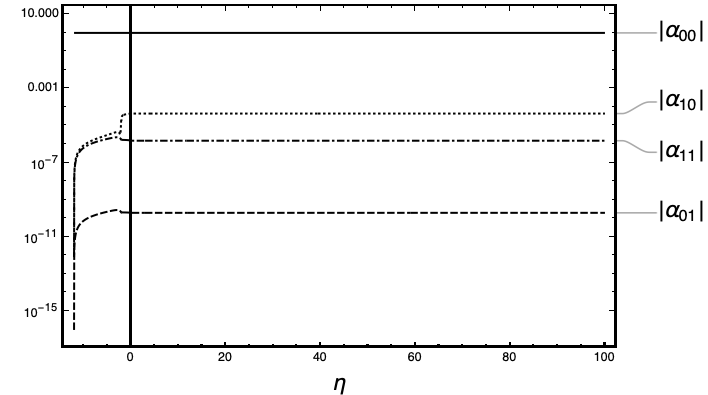}
 \caption{The evolution of $|\alpha_{00}|$, $|\alpha_{01}|$, $|\alpha_{10}|$ and $|\alpha_{11}|$. The value of $|\alpha_{00}|$ hardly changes as the produced probability amplitude for the other universe $\alpha_{11}$ as well as the entanglement amplitudes  $\alpha_{01}$, $\alpha_{10}$ are very small. The numerical violation of unitarity is kept under control.}\label{fig5}
\end{figure}

We assume the ansatz \eqref{bistate} and set $\alpha_{00}=1$, $\alpha_{01}=0$, $\alpha_{10}=0$ and $\alpha_{11}=0$ initially. Thus, initially there is no virtual universe $|q_1,p_1 \rangle$ and the perturbations on $|q_0,p_0 \rangle$ are expected to satisfy the Born-Oppenheimer dynamics. As expected from the dynamics \eqref{generaldynamics2}, such an initial solution is unstable and the other universe with its internal perturbation immediately emerges and influences the structure formation on the observable branch $|q_0,p_0 \rangle$. As a result, we obtain the dynamics of $\alpha_{00}$, $\alpha_{01}$, $\alpha_{10}$, and $\alpha_{11}$ as plotted in Fig. \ref{fig5}.

As can be seen in Fig.~\ref{fig5}, the amplitude $\alpha_{00}$
is only very mildly modified, while the other $\alpha_{ij}$ 
are only slightly increased, with a hierarchy of values. This
is a consequence of our chosen set of parameter values, and
a full study over the parameter space and for many different
wavelengths should be performed to evaluate its genericity.

\begin{figure}[t!]
    \centering
   \includegraphics[width=0.5\textwidth]{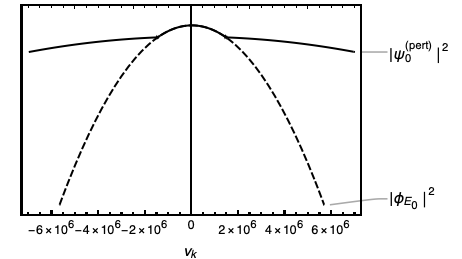}\\ \vspace{0.3cm}
   \includegraphics[width=0.5\textwidth]{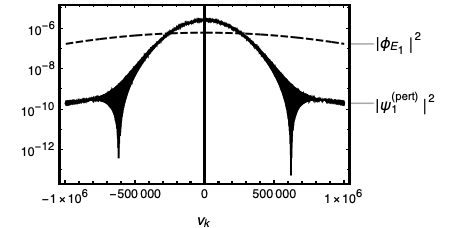}
 \caption{The probability distributions for the perturbation amplitude at $\eta=25$ when the perturbations are superhorizon and constant. {\it Top:} The Born-Oppenheimer perturbation state on the background ``0'' evolves into a combination with a small but discernible contribution from the Born-Oppenheimer perturbation state originating from the background ``1''. This contribution can be seen in the final distribution at very large amplitudes $v_{\bm{k}}$. {\it Bottom:} For background ``1'', the initial perturbation is eventually overcome by a dominant contribution from the perturbation associated with the background ``0'', leading to a significant shift in the probability distribution toward smaller values of $v_{\bm{k}}$.} \label{fig6}
\end{figure}

Figure~\ref{fig6} illustrates the changes observed in the initial perturbation states within both backgrounds when their interaction is considered. The probability distribution of the primordial amplitude is much more concentrated in the observed background than in the virtual one, indicating a larger amplification of amplitude in the latter. As a result of the full dynamics, the probability distribution of the perturbation amplitude in the first background exhibits a long tail at larger values of $v$, however, with relatively insignificant total probability. On the other hand, the distribution of the second perturbation becomes significantly altered, noticeably shifting toward the smaller values of $v$. These probability distributions evolve over time, and the plots represent a fixed moment of time at $\eta=25$. The new distributions exhibit visibly non-Gaussian characteristics, although symmetric (no skewness).

\section{Discussion}\label{Discussion}

We have developed a framework that fully captures the influence of the cosmological background on the linear perturbations. Choosing conveniently the Born-Oppenheimer universes for basis vectors in the state space, and truncating the model to 4 degrees of freedom, we have found that the expanding universe generically emerges in an entangled state. Upon projecting it onto an observable background, the primordial perturbation becomes well-defined but exhibits non-Gaussian behavior. However, the degree of non-Gaussianity appears to be very small in the numerical example presented herein. An analogous numerical computation (not shown in the paper) shows that reversing the roles of the two BO universes in the initial condition yields a similarly inverted final state.

While the amount of non-Gaussianity in our example is extremely small for the observable branch, it is premature to assert that the Born-Oppenheimer approximation will invariably yield such accurate results. The current study represents merely a proof of concept, i.e., the first step toward understanding the dynamics beyond the Born-Oppenheimer approximation. A more comprehensive numerical investigation should be constructed based on an analytical definition of the BO universes. This approach would enable one to manipulate parameters within the truncated model, such as the energies of backgrounds and the times of their bounces, to determine the situations in which calculating corrections to a Born-Oppenheimer universe is most useful. 

Moreover, we should include perturbation modes of other wavelengths in our analysis. Their presence may dramatically change the dynamical picture obtained thus far. Specifically, these modes could combine to exert a stronger influence on the backgrounds as well as become entangled with one another, yielding non-negligible effects on the expanding universe. Enhancing the method presented here by identifying the optimal basis or proposing an entirely different and more efficient approach could greatly help to derive new and important physical results.

\begin{acknowledgements}
P.M. acknowledges the support of the National Science Centre
(NCN, Poland) under the Research Grant No. 2018/30/E/ST2/00370.
\end{acknowledgements}

\appendix

\bibliography{references.bib}

\begin{thebibliography}{28}%
\makeatletter
\providecommand \@ifxundefined [1]{%
 \@ifx{#1\undefined}
}%
\providecommand \@ifnum [1]{%
 \ifnum #1\expandafter \@firstoftwo
 \else \expandafter \@secondoftwo
 \fi
}%
\providecommand \@ifx [1]{%
 \ifx #1\expandafter \@firstoftwo
 \else \expandafter \@secondoftwo
 \fi
}%
\providecommand \natexlab [1]{#1}%
\providecommand \enquote  [1]{``#1''}%
\providecommand \bibnamefont  [1]{#1}%
\providecommand \bibfnamefont [1]{#1}%
\providecommand \citenamefont [1]{#1}%
\providecommand \href@noop [0]{\@secondoftwo}%
\providecommand \href [0]{\begingroup \@sanitize@url \@href}%
\providecommand \@href[1]{\@@startlink{#1}\@@href}%
\providecommand \@@href[1]{\endgroup#1\@@endlink}%
\providecommand \@sanitize@url [0]{\catcode `\\12\catcode `\$12\catcode
  `\&12\catcode `\#12\catcode `\^12\catcode `\_12\catcode `\%12\relax}%
\providecommand \@@startlink[1]{}%
\providecommand \@@endlink[0]{}%
\providecommand \url  [0]{\begingroup\@sanitize@url \@url }%
\providecommand \@url [1]{\endgroup\@href {#1}{\urlprefix }}%
\providecommand \urlprefix  [0]{URL }%
\providecommand \Eprint [0]{\href }%
\providecommand \doibase [0]{http://dx.doi.org/}%
\providecommand \selectlanguage [0]{\@gobble}%
\providecommand \bibinfo  [0]{\@secondoftwo}%
\providecommand \bibfield  [0]{\@secondoftwo}%
\providecommand \translation [1]{[#1]}%
\providecommand \BibitemOpen [0]{}%
\providecommand \bibitemStop [0]{}%
\providecommand \bibitemNoStop [0]{.\EOS\space}%
\providecommand \EOS [0]{\spacefactor3000\relax}%
\providecommand \BibitemShut  [1]{\csname bibitem#1\endcsname}%
\let\auto@bib@innerbib\@empty
\bibitem [{\citenamefont {Guth}(1981)}]{Guth:1980zm}%
  \BibitemOpen
  \bibfield  {author} {\bibinfo {author} {\bibfnamefont {Alan~H.}\ \bibnamefont
  {Guth}},\ }\bibfield  {title} {\enquote {\bibinfo {title} {{The Inflationary
  Universe: A Possible Solution to the Horizon and Flatness Problems}},}\
  }\href {\doibase 10.1103/PhysRevD.23.347} {\bibfield  {journal} {\bibinfo
  {journal} {Phys. Rev. D}\ }\textbf {\bibinfo {volume} {23}},\ \bibinfo
  {pages} {347--356} (\bibinfo {year} {1981})}\BibitemShut {NoStop}%
\bibitem [{\citenamefont {Mukhanov}(2005)}]{Mukhanov:2005sc}%
  \BibitemOpen
  \bibfield  {author} {\bibinfo {author} {\bibfnamefont {V.}~\bibnamefont
  {Mukhanov}},\ }\href {\doibase 10.1017/CBO9780511790553} {\emph {\bibinfo
  {title} {{Physical Foundations of Cosmology}}}}\ (\bibinfo  {publisher}
  {Cambridge University Press},\ \bibinfo {address} {Oxford},\ \bibinfo {year}
  {2005})\BibitemShut {NoStop}%
\bibitem [{\citenamefont {Peter}\ and\ \citenamefont
  {Uzan}(2013)}]{Peter:2013avv}%
  \BibitemOpen
  \bibfield  {author} {\bibinfo {author} {\bibfnamefont {Patrick}\ \bibnamefont
  {Peter}}\ and\ \bibinfo {author} {\bibfnamefont {Jean-Philippe}\ \bibnamefont
  {Uzan}},\ }\href@noop {} {\emph {\bibinfo {title} {{Primordial
  Cosmology}}}},\ Oxford Graduate Texts\ (\bibinfo  {publisher} {Oxford
  University Press},\ \bibinfo {year} {2013})\BibitemShut {NoStop}%
\bibitem [{\citenamefont {Martin}\ \emph
  {et~al.}(2024{\natexlab{a}})\citenamefont {Martin}, \citenamefont
  {Ringeval},\ and\ \citenamefont {Vennin}}]{Martin:2024qnn}%
  \BibitemOpen
  \bibfield  {author} {\bibinfo {author} {\bibfnamefont {Jerome}\ \bibnamefont
  {Martin}}, \bibinfo {author} {\bibfnamefont {Christophe}\ \bibnamefont
  {Ringeval}}, \ and\ \bibinfo {author} {\bibfnamefont {Vincent}\ \bibnamefont
  {Vennin}},\ }\bibfield  {title} {\enquote {\bibinfo {title} {{Cosmic
  Inflation at the crossroads}},}\ }\href {\doibase
  10.1088/1475-7516/2024/07/087} {\bibfield  {journal} {\bibinfo  {journal}
  {JCAP}\ }\textbf {\bibinfo {volume} {07}},\ \bibinfo {pages} {087} (\bibinfo
  {year} {2024}{\natexlab{a}})},\ \Eprint {http://arxiv.org/abs/2404.10647}
  {arXiv:2404.10647 [astro-ph.CO]} \BibitemShut {NoStop}%
\bibitem [{\citenamefont {Borde}\ \emph {et~al.}(2003)\citenamefont {Borde},
  \citenamefont {Guth},\ and\ \citenamefont {Vilenkin}}]{Borde:2001nh}%
  \BibitemOpen
  \bibfield  {author} {\bibinfo {author} {\bibfnamefont {Arvind}\ \bibnamefont
  {Borde}}, \bibinfo {author} {\bibfnamefont {Alan~H.}\ \bibnamefont {Guth}}, \
  and\ \bibinfo {author} {\bibfnamefont {Alexander}\ \bibnamefont {Vilenkin}},\
  }\bibfield  {title} {\enquote {\bibinfo {title} {{Inflationary space-times
  are incomplete in past directions}},}\ }\href {\doibase
  10.1103/PhysRevLett.90.151301} {\bibfield  {journal} {\bibinfo  {journal}
  {Phys. Rev. Lett.}\ }\textbf {\bibinfo {volume} {90}},\ \bibinfo {pages}
  {151301} (\bibinfo {year} {2003})},\ \Eprint
  {http://arxiv.org/abs/gr-qc/0110012} {arXiv:gr-qc/0110012} \BibitemShut
  {NoStop}%
\bibitem [{\citenamefont {Lesnefsky}\ \emph {et~al.}(2023)\citenamefont
  {Lesnefsky}, \citenamefont {Easson},\ and\ \citenamefont
  {Davies}}]{Lesnefsky:2022fen}%
  \BibitemOpen
  \bibfield  {author} {\bibinfo {author} {\bibfnamefont {J.~E.}\ \bibnamefont
  {Lesnefsky}}, \bibinfo {author} {\bibfnamefont {D.~A.}\ \bibnamefont
  {Easson}}, \ and\ \bibinfo {author} {\bibfnamefont {P.~C.~W.}\ \bibnamefont
  {Davies}},\ }\bibfield  {title} {\enquote {\bibinfo {title}
  {{Past-completeness of inflationary spacetimes}},}\ }\href {\doibase
  10.1103/PhysRevD.107.044024} {\bibfield  {journal} {\bibinfo  {journal}
  {Phys. Rev. D}\ }\textbf {\bibinfo {volume} {107}},\ \bibinfo {pages}
  {044024} (\bibinfo {year} {2023})},\ \Eprint
  {http://arxiv.org/abs/2207.00955} {arXiv:2207.00955 [gr-qc]} \BibitemShut
  {NoStop}%
\bibitem [{\citenamefont {Peter}\ \emph {et~al.}(2006)\citenamefont {Peter},
  \citenamefont {Pinho},\ and\ \citenamefont {Pinto-Neto}}]{Peter_2006}%
  \BibitemOpen
  \bibfield  {author} {\bibinfo {author} {\bibfnamefont {Patrick}\ \bibnamefont
  {Peter}}, \bibinfo {author} {\bibfnamefont {Emanuel J.~C.}\ \bibnamefont
  {Pinho}}, \ and\ \bibinfo {author} {\bibfnamefont {Nelson}\ \bibnamefont
  {Pinto-Neto}},\ }\bibfield  {title} {\enquote {\bibinfo {title}
  {Gravitational wave background in perfect fluid quantum cosmologies},}\
  }\href {\doibase 10.1103/physrevd.73.104017} {\bibfield  {journal} {\bibinfo
  {journal} {Phys. Rev. D}\ }\textbf {\bibinfo {volume} {73}} (\bibinfo {year}
  {2006}),\ 10.1103/physrevd.73.104017}\BibitemShut {NoStop}%
\bibitem [{\citenamefont {Pinho}\ and\ \citenamefont
  {Pinto-Neto}(2007{\natexlab{a}})}]{Pinho_2007}%
  \BibitemOpen
  \bibfield  {author} {\bibinfo {author} {\bibfnamefont {Emanuel J.~C.}\
  \bibnamefont {Pinho}}\ and\ \bibinfo {author} {\bibfnamefont {Nelson}\
  \bibnamefont {Pinto-Neto}},\ }\bibfield  {title} {\enquote {\bibinfo {title}
  {Scalar and vector perturbations in quantum cosmological backgrounds},}\
  }\href {\doibase 10.1103/physrevd.76.023506} {\bibfield  {journal} {\bibinfo
  {journal} {Phys. Rev. D}\ }\textbf {\bibinfo {volume} {76}} (\bibinfo {year}
  {2007}{\natexlab{a}}),\ 10.1103/physrevd.76.023506}\BibitemShut {NoStop}%
\bibitem [{\citenamefont {Ashtekar}\ \emph {et~al.}(2009)\citenamefont
  {Ashtekar}, \citenamefont {Kaminski},\ and\ \citenamefont
  {Lewandowski}}]{PhysRevD.79.064030}%
  \BibitemOpen
  \bibfield  {author} {\bibinfo {author} {\bibfnamefont {Abhay}\ \bibnamefont
  {Ashtekar}}, \bibinfo {author} {\bibfnamefont {Wojciech}\ \bibnamefont
  {Kaminski}}, \ and\ \bibinfo {author} {\bibfnamefont {Jerzy}\ \bibnamefont
  {Lewandowski}},\ }\bibfield  {title} {\enquote {\bibinfo {title} {Quantum
  field theory on a cosmological, quantum space-time},}\ }\href {\doibase
  10.1103/PhysRevD.79.064030} {\bibfield  {journal} {\bibinfo  {journal} {Phys.
  Rev. D}\ }\textbf {\bibinfo {volume} {79}},\ \bibinfo {pages} {064030}
  (\bibinfo {year} {2009})}\BibitemShut {NoStop}%
\bibitem [{\citenamefont {Ma\l{}kiewicz}\ and\ \citenamefont
  {Miroszewski}(2021)}]{Ma_kiewicz_2021}%
  \BibitemOpen
  \bibfield  {author} {\bibinfo {author} {\bibfnamefont {Przemys\l{}aw}\
  \bibnamefont {Ma\l{}kiewicz}}\ and\ \bibinfo {author} {\bibfnamefont {Artur}\
  \bibnamefont {Miroszewski}},\ }\bibfield  {title} {\enquote {\bibinfo {title}
  {Dynamics of primordial fields in quantum cosmological spacetimes},}\ }\href
  {\doibase 10.1103/physrevd.103.083529} {\bibfield  {journal} {\bibinfo
  {journal} {Phys. Rev. D}\ }\textbf {\bibinfo {volume} {103}} (\bibinfo {year}
  {2021}),\ 10.1103/physrevd.103.083529}\BibitemShut {NoStop}%
\bibitem [{\citenamefont {Kiefer}(2012)}]{Kiefer:2004xyv}%
  \BibitemOpen
  \bibfield  {author} {\bibinfo {author} {\bibfnamefont {Claus}\ \bibnamefont
  {Kiefer}},\ }\href@noop {} {\emph {\bibinfo {title} {{Quantum gravity}}}},\
  \bibinfo {edition} {3rd}\ ed.,\ Vol.\ \bibinfo {volume} {155}\ (\bibinfo
  {publisher} {Clarendon},\ \bibinfo {address} {Oxford},\ \bibinfo {year}
  {2012})\BibitemShut {NoStop}%
\bibitem [{\citenamefont {Bini}\ \emph {et~al.}(2013)\citenamefont {Bini},
  \citenamefont {Esposito}, \citenamefont {Kiefer}, \citenamefont {Kr\"amer},\
  and\ \citenamefont {Pessina}}]{PhysRevD.87.104008}%
  \BibitemOpen
  \bibfield  {author} {\bibinfo {author} {\bibfnamefont {Donato}\ \bibnamefont
  {Bini}}, \bibinfo {author} {\bibfnamefont {Giampiero}\ \bibnamefont
  {Esposito}}, \bibinfo {author} {\bibfnamefont {Claus}\ \bibnamefont
  {Kiefer}}, \bibinfo {author} {\bibfnamefont {Manuel}\ \bibnamefont
  {Kr\"amer}}, \ and\ \bibinfo {author} {\bibfnamefont {Francesco}\
  \bibnamefont {Pessina}},\ }\bibfield  {title} {\enquote {\bibinfo {title} {On
  the modification of the cosmic microwave background anisotropy spectrum from
  canonical quantum gravity},}\ }\href {\doibase 10.1103/PhysRevD.87.104008}
  {\bibfield  {journal} {\bibinfo  {journal} {Phys. Rev. D}\ }\textbf {\bibinfo
  {volume} {87}},\ \bibinfo {pages} {104008} (\bibinfo {year}
  {2013})}\BibitemShut {NoStop}%
\bibitem [{\citenamefont {Chataignier}\ and\ \citenamefont
  {Kr\"amer}(2021)}]{PhysRevD.103.066005}%
  \BibitemOpen
  \bibfield  {author} {\bibinfo {author} {\bibfnamefont {Leonardo}\
  \bibnamefont {Chataignier}}\ and\ \bibinfo {author} {\bibfnamefont {Manuel}\
  \bibnamefont {Kr\"amer}},\ }\bibfield  {title} {\enquote {\bibinfo {title}
  {Unitarity of quantum-gravitational corrections to primordial fluctuations in
  the born-oppenheimer approach},}\ }\href {\doibase
  10.1103/PhysRevD.103.066005} {\bibfield  {journal} {\bibinfo  {journal}
  {Phys. Rev. D}\ }\textbf {\bibinfo {volume} {103}},\ \bibinfo {pages}
  {066005} (\bibinfo {year} {2021})}\BibitemShut {NoStop}%
\bibitem [{\citenamefont {Kamenshchik}\ \emph {et~al.}(2013)\citenamefont
  {Kamenshchik}, \citenamefont {Tronconi},\ and\ \citenamefont
  {Venturi}}]{KAMENSHCHIK2013518}%
  \BibitemOpen
  \bibfield  {author} {\bibinfo {author} {\bibfnamefont {Alexander~Y.}\
  \bibnamefont {Kamenshchik}}, \bibinfo {author} {\bibfnamefont {Alessandro}\
  \bibnamefont {Tronconi}}, \ and\ \bibinfo {author} {\bibfnamefont {Giovanni}\
  \bibnamefont {Venturi}},\ }\bibfield  {title} {\enquote {\bibinfo {title}
  {Inflation and quantum gravity in a born-oppenheimer context},}\ }\href
  {\doibase https://doi.org/10.1016/j.physletb.2013.08.067} {\bibfield
  {journal} {\bibinfo  {journal} {Phys. Lett. B}\ }\textbf {\bibinfo {volume}
  {726}},\ \bibinfo {pages} {518--522} (\bibinfo {year} {2013})}\BibitemShut
  {NoStop}%
\bibitem [{\citenamefont {Gomar}\ \emph {et~al.}(2014)\citenamefont {Gomar},
  \citenamefont {Fern\'andez-M\'endez}, \citenamefont {Marug\'an},\ and\
  \citenamefont {Olmedo}}]{Gomar_2014}%
  \BibitemOpen
  \bibfield  {author} {\bibinfo {author} {\bibfnamefont {Laura~Castell\'o}\
  \bibnamefont {Gomar}}, \bibinfo {author} {\bibfnamefont {Mikel}\ \bibnamefont
  {Fern\'andez-M\'endez}}, \bibinfo {author} {\bibfnamefont {Guillermo
  A.~Mena}\ \bibnamefont {Marug\'an}}, \ and\ \bibinfo {author} {\bibfnamefont
  {Javier}\ \bibnamefont {Olmedo}},\ }\bibfield  {title} {\enquote {\bibinfo
  {title} {Cosmological perturbations in hybrid loop quantum cosmology:
  Mukhanov-sasaki variables},}\ }\href {\doibase 10.1103/physrevd.90.064015}
  {\bibfield  {journal} {\bibinfo  {journal} {Phys. Rev. D}\ }\textbf {\bibinfo
  {volume} {90}} (\bibinfo {year} {2014}),\
  10.1103/physrevd.90.064015}\BibitemShut {NoStop}%
\bibitem [{\citenamefont {Maniccia}\ and\ \citenamefont
  {Montani}(2022)}]{PhysRevD.105.086014}%
  \BibitemOpen
  \bibfield  {author} {\bibinfo {author} {\bibfnamefont {Giulia}\ \bibnamefont
  {Maniccia}}\ and\ \bibinfo {author} {\bibfnamefont {Giovanni}\ \bibnamefont
  {Montani}},\ }\bibfield  {title} {\enquote {\bibinfo {title} {Quantum gravity
  corrections to the matter dynamics in the presence of a reference fluid},}\
  }\href {\doibase 10.1103/PhysRevD.105.086014} {\bibfield  {journal} {\bibinfo
   {journal} {Phys. Rev. D}\ }\textbf {\bibinfo {volume} {105}},\ \bibinfo
  {pages} {086014} (\bibinfo {year} {2022})}\BibitemShut {NoStop}%
\bibitem [{\citenamefont {Born}\ and\ \citenamefont
  {Oppenheimer}(1927)}]{Born:1927rpw}%
  \BibitemOpen
  \bibfield  {author} {\bibinfo {author} {\bibfnamefont {M.}~\bibnamefont
  {Born}}\ and\ \bibinfo {author} {\bibfnamefont {R.}~\bibnamefont
  {Oppenheimer}},\ }\bibfield  {title} {\enquote {\bibinfo {title} {{Zur
  Quantentheorie der Molekeln}},}\ }\href {\doibase 10.1002/andp.19273892002}
  {\bibfield  {journal} {\bibinfo  {journal} {Annalen Phys.}\ }\textbf
  {\bibinfo {volume} {389}},\ \bibinfo {pages} {457--484} (\bibinfo {year}
  {1927})}\BibitemShut {NoStop}%
\bibitem [{\citenamefont {Mott}(1931)}]{Mott:1931}%
  \BibitemOpen
  \bibfield  {author} {\bibinfo {author} {\bibfnamefont {N.~F.}\ \bibnamefont
  {Mott}},\ }\bibfield  {title} {\enquote {\bibinfo {title} {{On the theory of
  excitation by collision with heavy particles}},}\ }\href@noop {} {\bibfield
  {journal} {\bibinfo  {journal} {Proc. Cambridge Phil. Soc.}\ }\textbf
  {\bibinfo {volume} {27}},\ \bibinfo {pages} {553--560} (\bibinfo {year}
  {1931})}\BibitemShut {NoStop}%
\bibitem [{\citenamefont {Mukhanov}\ \emph {et~al.}(1992)\citenamefont
  {Mukhanov}, \citenamefont {Feldman},\ and\ \citenamefont
  {Brandenberger}}]{Mukhanov:1990me}%
  \BibitemOpen
  \bibfield  {author} {\bibinfo {author} {\bibfnamefont {Viatcheslav~F.}\
  \bibnamefont {Mukhanov}}, \bibinfo {author} {\bibfnamefont {H.~A.}\
  \bibnamefont {Feldman}}, \ and\ \bibinfo {author} {\bibfnamefont {Robert~H.}\
  \bibnamefont {Brandenberger}},\ }\bibfield  {title} {\enquote {\bibinfo
  {title} {{Theory of cosmological perturbations. Part 1. Classical
  perturbations. Part 2. Quantum theory of perturbations. Part 3.
  Extensions}},}\ }\href {\doibase 10.1016/0370-1573(92)90044-Z} {\bibfield
  {journal} {\bibinfo  {journal} {Phys. Rept.}\ }\textbf {\bibinfo {volume}
  {215}},\ \bibinfo {pages} {203--333} (\bibinfo {year} {1992})}\BibitemShut
  {NoStop}%
\bibitem [{\citenamefont {Pinho}\ and\ \citenamefont
  {Pinto-Neto}(2007{\natexlab{b}})}]{Pinho:2006ym}%
  \BibitemOpen
  \bibfield  {author} {\bibinfo {author} {\bibfnamefont {Emanuel J.~C.}\
  \bibnamefont {Pinho}}\ and\ \bibinfo {author} {\bibfnamefont {Nelson}\
  \bibnamefont {Pinto-Neto}},\ }\bibfield  {title} {\enquote {\bibinfo {title}
  {{Scalar and vector perturbations in quantum cosmological backgrounds}},}\
  }\href {\doibase 10.1103/PhysRevD.76.023506} {\bibfield  {journal} {\bibinfo
  {journal} {Phys. Rev. D}\ }\textbf {\bibinfo {volume} {76}},\ \bibinfo
  {pages} {023506} (\bibinfo {year} {2007}{\natexlab{b}})},\ \Eprint
  {http://arxiv.org/abs/hep-th/0610192} {arXiv:hep-th/0610192} \BibitemShut
  {NoStop}%
\bibitem [{\citenamefont {Ma\l{}kiewicz}(2019)}]{Ma_kiewicz_2019}%
  \BibitemOpen
  \bibfield  {author} {\bibinfo {author} {\bibfnamefont {Przemys\l{}aw}\
  \bibnamefont {Ma\l{}kiewicz}},\ }\bibfield  {title} {\enquote {\bibinfo
  {title} {Hamiltonian formalism and gauge-fixing conditions for cosmological
  perturbation theory},}\ }\href {\doibase 10.1088/1361-6382/ab45aa} {\bibfield
   {journal} {\bibinfo  {journal} {Class. Quantum Grav.}\ }\textbf {\bibinfo
  {volume} {36}},\ \bibinfo {pages} {215003} (\bibinfo {year}
  {2019})}\BibitemShut {NoStop}%
\bibitem [{\citenamefont {Martin}\ \emph {et~al.}(2022)\citenamefont {Martin},
  \citenamefont {Ma\l{}kiewicz},\ and\ \citenamefont {Peter}}]{Martin_2022}%
  \BibitemOpen
  \bibfield  {author} {\bibinfo {author} {\bibfnamefont {Jaime de~Cabo}\
  \bibnamefont {Martin}}, \bibinfo {author} {\bibfnamefont {Przemys\l{}aw}\
  \bibnamefont {Ma\l{}kiewicz}}, \ and\ \bibinfo {author} {\bibfnamefont
  {Patrick}\ \bibnamefont {Peter}},\ }\bibfield  {title} {\enquote {\bibinfo
  {title} {Unitarily inequivalent quantum cosmological bouncing models},}\
  }\href {\doibase 10.1103/physrevd.105.023522} {\bibfield  {journal} {\bibinfo
   {journal} {Phys. Rev. D}\ }\textbf {\bibinfo {volume} {105}} (\bibinfo
  {year} {2022}),\ 10.1103/physrevd.105.023522}\BibitemShut {NoStop}%
\bibitem [{\citenamefont {Martin}\ \emph
  {et~al.}(2024{\natexlab{b}})\citenamefont {Martin}, \citenamefont
  {Ma\l{}kiewicz},\ and\ \citenamefont {Peter}}]{Martin:2022ptk}%
  \BibitemOpen
  \bibfield  {author} {\bibinfo {author} {\bibfnamefont {Jaime de~Cabo}\
  \bibnamefont {Martin}}, \bibinfo {author} {\bibfnamefont {Przemys\l{}aw}\
  \bibnamefont {Ma\l{}kiewicz}}, \ and\ \bibinfo {author} {\bibfnamefont
  {Patrick}\ \bibnamefont {Peter}},\ }\bibfield  {title} {\enquote {\bibinfo
  {title} {{Ambiguous power spectrum from a quantum bounce}},}\ }\href
  {\doibase 10.1103/PhysRevD.109.066009} {\bibfield  {journal} {\bibinfo
  {journal} {Phys. Rev. D}\ }\textbf {\bibinfo {volume} {109}},\ \bibinfo
  {pages} {066009} (\bibinfo {year} {2024}{\natexlab{b}})},\ \Eprint
  {http://arxiv.org/abs/2212.12484} {arXiv:2212.12484 [gr-qc]} \BibitemShut
  {NoStop}%
\bibitem [{\citenamefont {Klauder}(2015)}]{Klauder:2015ifa}%
  \BibitemOpen
  \bibfield  {author} {\bibinfo {author} {\bibfnamefont {John~R.}\ \bibnamefont
  {Klauder}},\ }\href {\doibase 10.1142/9452} {\emph {\bibinfo {title}
  {{Enhanced quantization}: {Particles, fields and gravity}}}}\ (\bibinfo
  {publisher} {World Scientific},\ \bibinfo {address} {Hackensack},\ \bibinfo
  {year} {2015})\BibitemShut {NoStop}%
\bibitem [{\citenamefont {Gazeau}\ and\ \citenamefont
  {Murenzi}(2016)}]{Gazeau:2015nkc}%
  \BibitemOpen
  \bibfield  {author} {\bibinfo {author} {\bibfnamefont {Jean~Pierre}\
  \bibnamefont {Gazeau}}\ and\ \bibinfo {author} {\bibfnamefont {Romain}\
  \bibnamefont {Murenzi}},\ }\bibfield  {title} {\enquote {\bibinfo {title}
  {{Covariant affine integral quantization(s)}},}\ }\href {\doibase
  10.1063/1.4949366} {\bibfield  {journal} {\bibinfo  {journal} {J. Math.
  Phys.}\ }\textbf {\bibinfo {volume} {57}},\ \bibinfo {pages} {052102--1}
  (\bibinfo {year} {2016})},\ \Eprint {http://arxiv.org/abs/1512.08274}
  {arXiv:1512.08274 [quant-ph]} \BibitemShut {NoStop}%
\bibitem [{\citenamefont {Ma\l{}kiewicz}\ \emph {et~al.}(2020)\citenamefont
  {Ma\l{}kiewicz}, \citenamefont {Peter},\ and\ \citenamefont
  {Vitenti}}]{Ma_kiewicz_2020}%
  \BibitemOpen
  \bibfield  {author} {\bibinfo {author} {\bibfnamefont {Przemys\l{}aw}\
  \bibnamefont {Ma\l{}kiewicz}}, \bibinfo {author} {\bibfnamefont {Patrick}\
  \bibnamefont {Peter}}, \ and\ \bibinfo {author} {\bibfnamefont {S.~D.~P.}\
  \bibnamefont {Vitenti}},\ }\bibfield  {title} {\enquote {\bibinfo {title}
  {Quantum empty bianchi i spacetime with internal time},}\ }\href {\doibase
  10.1103/physrevd.101.046012} {\bibfield  {journal} {\bibinfo  {journal}
  {Phys. Rev. D}\ }\textbf {\bibinfo {volume} {101}} (\bibinfo {year} {2020}),\
  10.1103/physrevd.101.046012}\BibitemShut {NoStop}%
\bibitem [{\citenamefont {Bergeron}\ \emph {et~al.}(2024)\citenamefont
  {Bergeron}, \citenamefont {Gazeau}, \citenamefont {Ma\l{}kiewicz},\ and\
  \citenamefont {Peter}}]{Bergeron:2023zzo}%
  \BibitemOpen
  \bibfield  {author} {\bibinfo {author} {\bibfnamefont {Herv\'e}\ \bibnamefont
  {Bergeron}}, \bibinfo {author} {\bibfnamefont {Jean-Pierre}\ \bibnamefont
  {Gazeau}}, \bibinfo {author} {\bibfnamefont {Przemys\l{}aw}\ \bibnamefont
  {Ma\l{}kiewicz}}, \ and\ \bibinfo {author} {\bibfnamefont {Patrick}\
  \bibnamefont {Peter}},\ }\bibfield  {title} {\enquote {\bibinfo {title} {{New
  class of exact coherent states: Enhanced quantization of motion on the half
  line}},}\ }\href {\doibase 10.1103/PhysRevD.109.023516} {\bibfield  {journal}
  {\bibinfo  {journal} {Phys. Rev. D}\ }\textbf {\bibinfo {volume} {109}},\
  \bibinfo {pages} {023516} (\bibinfo {year} {2024})},\ \Eprint
  {http://arxiv.org/abs/2310.16868} {arXiv:2310.16868 [quant-ph]} \BibitemShut
  {NoStop}%
\bibitem [{\citenamefont {Caderni}\ and\ \citenamefont
  {Martellini}(1984)}]{Caderni:1984pw}%
  \BibitemOpen
  \bibfield  {author} {\bibinfo {author} {\bibfnamefont {N.}~\bibnamefont
  {Caderni}}\ and\ \bibinfo {author} {\bibfnamefont {M.}~\bibnamefont
  {Martellini}},\ }\bibfield  {title} {\enquote {\bibinfo {title} {{Third
  quantization formalism for Hamiltonian cosmologies}},}\ }\href {\doibase
  10.1007/BF02080689} {\bibfield  {journal} {\bibinfo  {journal} {Int. J.
  Theor. Phys.}\ }\textbf {\bibinfo {volume} {23}},\ \bibinfo {pages}
  {233--249} (\bibinfo {year} {1984})}\BibitemShut {NoStop}%
\end{thebibliography}%

\end{document}